\documentclass{ifacconf}

\usepackage{graphicx}      
\usepackage{natbib}        
\usepackage{amsmath}
\usepackage{placeins}
\usepackage{caption} 
\usepackage{subcaption} 
\usepackage{amssymb}
\usepackage{color}
\usepackage{amsfonts}
\usepackage{mathtools}

\usepackage{theoremref}
\usepackage[algo2e, ruled]{algorithm2e} 
\usepackage{commath}
\usepackage[utf8]{inputenc} 

\begin{document}

\begin{frontmatter}

\title{Control Synthesis for Multi-Agent Systems under Metric Interval Temporal Logic Specifications\thanksref{footnoteinfo}} 

\thanks[footnoteinfo]{This work was supported by the H2020 ERC Starting Grand BUCOPHSYS, the Swedish Research Council (VR), the Swedish Foundation for Strategic Research (SSF) and the Knut och Alice Wallenberg Foundation.}

\author[First]{Sofie Andersson} 
\author[First]{Alexandros Nikou} 
\author[First]{Dimos V. Dimarogonas}

\address[First]{ ACCESS Linnaeus Center, School of Electrical Engineering and KTH Center for Autonomous Systems, KTH Royal Institute of Technology, SE-100 44, Stockholm, Sweden. \\ E-mail: \{sofa, anikou, dimos\}@kth.se}

\begin{abstract}                
This paper presents a framework for automatic synthesis of a control sequence for multi-agent systems governed by continuous linear dynamics under timed constraints. First, the motion of the agents in the workspace is abstracted into individual Transition Systems (TS). Second, each agent is assigned with an individual formula given in Metric Interval Temporal Logic (MITL) and in parallel, the team of agents is assigned with a collaborative team formula. The proposed method is based on a correct-by-construction control synthesis method, and hence guarantees that the resulting closed-loop system will satisfy the specifications. The specifications considers boolean-valued properties under real-time. Extended simulations has been performed in order to demonstrate the efficiency of the proposed controllers.
\end{abstract}


\end{frontmatter}

\section{Introduction} \label{sec:intro}

Multi-agent systems are composed by $N\geq 2$ number of agents which interact in an environment. Cooperative control for multi-agent systems allows the agents to collaborate on tasks and plan more efficiently. In this paper, the former is considered by regarding collaborative team specifications which requires more than one agent to satisfy some property at the same time. The aim is to construct a framework that will start from an environment and a set of tasks, both local (i.e. specific to an individual agent) and global (i.e. requires collaboration between multiple agents), and yield the closed-loop system that will achieve satisfaction of the specifications, by control synthesis.
 
The specification language that has been introduced to express such tasks is Linear Temporal Logic (LTL) (see e.g., \citep{loizou_2004}). The general framework that is used is based on a three-steps procedure (\citep{A1, LTL1}): First the agent dynamics is abstracted into a Transition System. Second a discrete plan that meets the high level task is synthesized. Third, this plan is translated into a sequence of continuous controllers for the original system.

Control synthesis for multi-agent systems under LTL specifications has been addressed in \citep{belta_cdc_reduced_communication, guo_2015_reconfiguration, zavlanos_2016_multi-agent_LTL}. Due to the fact that we are interested in imposing timed constraints to the system, the aforementioned works cannot be directly utilized. Timed constraints have been introduced for the single agent case in \citep{A2, murray_2015_stl, MITL2TA, baras_MTL_2016} and for the multi-agent case in \citep{frazzoli_MTL, X2}. Authors in \citep{frazzoli_MTL} addressed the vehicle routing problem, under Metric Temporal Logic (MTL) specifications. The corresponding approach does not rely on automata-based verification, as it is based on a construction of linear inequalities and the solution of a resulting Mixed-Integer Linear Programming (MILP) problem. In our previous work \citep{X2}, we proposed an automatic framework for multi-agent systems such that each agent satisfies an individual formula and the team of agents one global formula. 

The approach to solution suggested in this paper follows similar principles as in \citep{X2}. Here however, we start from the continuous linear system itself rather than assuming an abstraction, by adding a way to abstract the environment in a suitable manner such that the transition time is taken explicitly into account. The suggested abstraction is based on the work presented in \citep{A2}, which considered time bounds on facet reachability for a continuous-time multi-affine single agent system. Here, we consider multi-agent systems and suggest an alternative time estimation and provide a proof for its validity. Furthermore, we present alternative definitions of the local BWTS, the product BWTS and the global BWTS, compared to the work presented in \citep{X2}. The definitions suggested here requires a smaller number of states and hence, a lower computational demand. The drawback of the suggested definitions is an increased risk of a false negative result and a required modification to the applied graph-search-algorithm. However, this will have no effect on the fact that the method is correct-by-construction. The method, in its entirety, has been implemented in simulations, demonstrating the satisfaction of the specifications through the resulting controller. 

The contribution of this paper is summarized in four parts; (1) it extends the method suggested in \citep{X2} with the ability to define the environment directly as a continuous linear system rather than treating the abstraction as a given, (2) it provides for a less computationally demanding alternative, (3) simulation results which support the claims are included, (4) it considers linear dynamics in contrast to the already investigated (in \citep{X2}) single integrator. 

This paper is structured as follows. Section \ref{sec:PaN} introduces some preliminaries and notations that will be applied throughout the paper, Section \ref{sec:prob} defines the considered problem and Section \ref{sec:sol} presents the main result, namely the solution framework. Finally, simulation result is presented in Section \ref{sec:ex}, illustrating the framework when applied to a simple example, and conclusions are made in section \ref{sec:conc}.

\section{Preliminaries and Notation}\label{sec:PaN}

In this section, the mathematical notation and preliminary definitions from formal methods that are required for this paper are introduced.

Given a set $S$, we denote by $|S|, 2^S$ its cardinality and the set of all its subsets respectively. Let $A \in \mathbb{R}^{n \times m}, B \in \mathbb{R}^n$ be a matrix and a vector respectively. Denote by $[A]_{ij}$ the element in the $i$-th row and $j$-th column of matrix $A$. Similarly, denote by $[B]_i$ the $i$-th element of vector $B$.

Given a set of nonnegative rational numbers $\mathbb{T}\subset \mathbb{Q}_+$ a time sequence is defined as:
\begin{defn}\thlabel{def:Tseq}\citep{TA1}
A \textit{time sequence} $\tau=\tau_0\tau_1...$ is an infinite sequence of time values which satisfies all the following:
\begin{itemize}
\item $\tau_i\in \mathbb{T}\subset\mathbb{Q}_+$,
\item $\tau_i<\tau_{i+1}, \: \forall i\geq 0$ and
\item $\exists i\geq 1, \: s.t. \: \tau_i>t, \: \forall t\in \mathbb{T}$.
\end{itemize}
\end{defn}

An \textit{atomic proposition} $ap$ is a statement over the system variables that is either true ($\top$) or false ($\perp$).


\begin{defn}\thlabel{def:WTS}
A \textit{Weighted Transition System} (WTS) is a tuple
$T=(\Pi, \Pi_{init},\Sigma,\rightarrow,AP,L,d)$
where 
\begin{itemize}
\item $\Pi=\{r_i : i=0,...,M\}$ is a set of states,
\item $\Pi_{init}\subset \Pi$ is a set of initial states,
\item $\Sigma=\{\sigma_i : i=0,...,l\}$ is a set of inputs,
\item $\rightarrow \subset \Pi \times \Sigma \times \Pi$ is a transition map; the expression $r_i \overset{\sigma_j}{\rightarrow} r_k$ is used to express transition from $r_i$ to $r_k$ under the action $\sigma_j$,
\item $AP$ is a set of observations (atomic propositions),
\item $L:\Pi \to 2^{AP}$ is an observation map and
\item $d:\rightarrow \to \mathbb{R}_+$ is a positive weight assignment map; the expression $d(r_i, \sigma_j, r_k)$ is used to express the weight assigned to the transition $r_i\overset{\sigma_j}{\rightarrow} r_k$.
\end{itemize}
\end{defn}

\begin{defn}\thlabel{def:Trun}
A \textit{timed run} $r^t=(r(0),\tau_0)(r(1),\tau_1)...$ of a WTS $T$ is an infinite sequence where $r(0)\in \Pi_{init}$, and  $r(j)\in \Pi$, $r(j) \overset{\sigma_i}{\rightarrow} r(j+1)$ $\forall j\geq 1$ s.t.
\begin{itemize}
\item $\tau_0=0,$
\item $\tau_{j+1}=\tau_j+d(r(j),\sigma_i, r(j+1)), \: \forall j \geq 1$,
\end{itemize}
for some $\sigma_i \in \Sigma$.
\end{defn}

\begin{defn}
A \textit{timed word} produced by a timed run is an infinite sequence of pairs \\$w(r^t)=(L(r(0)),\tau_0)(L(r(1)),\tau_1)...$, where \\$r^t=(r(0), \tau_0)(r(1),\tau_1)...$ is the timed run.
\end{defn}

\begin{defn}\thlabel{def:semMITL} 
The \textit{syntax of MITL} over a set of atomic propositions $AP$ is defined by the grammar
\begin{equation}
\label{eq:MITL}
\phi:=\top \;| \; ap\;|\; \neg \; \phi\;|\; \phi\vee \psi\;|\;\phi \: \mathcal{U}_{[a,b]}\:\psi
\end{equation}
where $ap\in AP$ and $\phi$, $\psi$ are formulas over $AP$. The operators are \textit{Negation} ($\neg$), \textit{Conjunction} ($\vee$) and \textit{Until} ($\mathcal{U}$) respectively. The extended operators \textit{Eventually} ($\lozenge$) and \textit{Always} ($\square$) are defined as:
\begin{subequations}
\begin{align}
\lozenge_{[a,b]} \phi &:= \top \mathcal{U}_{[a,b]} \phi, \\
\square_{[a,b]} \phi &:=\neg \lozenge_{[a,b]}\neg \phi.
\end{align}
\end{subequations}

 Given a timed run $r^t=(r(0),\tau_0)(r(1),\tau_1),...,(r(n),\tau_n)$ of a WTS, the semantics of the satisfaction relation is then defined as: 
\begin{align}
(r^t,i) &\models  ap \Leftrightarrow ap \in L(r(i)), \nonumber\\
(r^t,i) &\models \neg\phi \Leftrightarrow (r^t,i) \nvDash \phi, \nonumber\\
(r^t,i) &\models \phi\wedge\psi \Leftrightarrow (r^t,i) \models \phi \: \text{and} \: (r^t,i) \models \psi, \nonumber\\
(r^t,i) &\models  \phi\:\mathcal{U}_{[a, b]}\:\psi \Leftrightarrow  \exists j\in [a, b], \: s.t. \:(r^t,j) \models \psi\enspace \text{and} \nonumber\\
 &\hspace{+45mm}\forall i\leq j, (r^t,i)\vDash\phi. \notag
\end{align}
\end{defn}
\begin{defn}
A \textit{clock constraint} $\Phi_x$ is a conjunctive formula on the form $x\bowtie a$, where $\bowtie \in \{<, >, \leq, \geq\}$, $x$ is a clock and $a$ is some constant. Let $\Phi_\mathcal{X}$ denote the \textit{set of clock constraints}. 
\end{defn}
The TBA was first introduced in \citep{TA1} and is defined as
\begin{defn}\thlabel{def:TBA} 
A \textit{Timed Büchi Automaton} (TBA) is a tuple\\ $A=(S,S_0,\mathcal{X},I,E,F,AP, \mathcal{L})$ where
\begin{itemize}
\item $S=\{s_i : i=0,1,...,M\}$ is a finite set of locations,
\item $S_0\in S$ is the set of initial locations,
\item $\mathcal{X}$ is a finite set of clocks,
\item $I:S\rightarrow\Phi_\mathcal{X}$ is a map labelling each state $s_i$ with some clock constraints $\Phi_\mathcal{X}$,
\item $E\subseteq S\times \Phi_\mathcal{X}\times 2^\mathcal{X}\times S$ is a set of transitions and
\item $F\subset S$ is a set of accepting locations,
\item $AP$ is a finite set of atomic propositions,
\item $\mathcal{L}$ is a labelling function, labelling every state with a subset of atomic propositions.
\end{itemize}
A state of $A$ is a pair $(s,v)$ where $s\in S$ is a location and $v$ is a clock valuation that satisfies the clock constraint $I(s)$. The initial state of $A$ is a pair $(s_0,(0,0,...,0))$, where $s_0\in S_0$ and the null-vector $(0,0,...,0)$ is a vector of $\abs{\mathcal{X}}$ number of valuations $v_i=0$. For the semantics and examples of the above TBA definition we refer the reader to \citep{alex_acc_2017}.
\end{defn}

It has been shown in previous work \citep{MITL1} that any MITL formula can be algorithmically translated to a TBA such that the language that satisfies the MITL formula is also the language that produces accepting runs by the TBA. The TBA expresses all possibilities, both satisfaction and violation of the MITL formula. All timed runs which result in the satisfaction of the MITL formula are called accepting:
\begin{defn}\thlabel{def:Arun}
An \textit{accepting run} is a run for which there are infinitely many $j\geq 0$ s.t. $q_j\in F$, i.e. a run which consists of infinitely many accepting states.
\end{defn}

In motion-planning, the movement of an agent can be described by a timed run. For the multi-agent case, the movement of all agents can be collectively described by a collective run. The definition is 
\begin{defn}\label{def:Grun}\citep{X2}
The \textit{collective timed run} $r_G=(r_G(0), \tau_G(0))(r_G(1),\tau_G(1))...$ of $N$ agents, is defined as follows
\begin{itemize}
\item $(r_G(0), \tau_G(0))=(r_1(0),...,r_N(0),\tau_G(0))$
\item $(r_G(i+1),\tau_G(i+1))=(r_1(j_1),...,r_N(j_N),\tau_G(i+1))$, for $i\geq 0$ where $(r_G(i),\tau_G(i))=(r_1(i),...,r_N(i),\tau_G(i))$ and
\begin{itemize}
\item $l=\underset{k\in I}{\text{argmin}}\{\tau_k(i_k+1)\}$,
\item $\tau_G(i+1)=\tau_l(i_l+1)$,
\item $r_k(j_k)=\left\{ \begin{array}{ll}
r_l(i_l+1),&\text{if } k=l\\
r_k(i_l),&\text{otherwise.} 
\end{array}\right.$
\end{itemize}
\end{itemize}
\end{defn}

\section{Problem Definition}\label{sec:prob}
\subsection{System Model}
Consider $N$ agents performing in a bounded workspace $X \subset \mathbb{R}^n$ and governed by the dynamics 
\begin{align}\label{eq:multi-agent dyn}
\dot{x}_m &= A_mx_m+B_mu_m,\: \: m\in \mathcal{I}, \notag \\
x_m(0) &= x_m^0,  x_m\in X
\end{align}
where $\mathcal{I} = \{1,...,N\}$ is a set containing a label for each agent.

\subsection{Problem Statement}
The problem considered in this paper consists in synthesizing a control input sequence, $u_m, m \in \mathcal{I}$, such that each agent satisfies a local individual MITL formula $\phi_m$ over the set of atomic propositions $AP_m$. At the same time, the team of agents should satisfy a team specification MITL formula $\phi_G$ over the set of atomic propositions $AP_G=\underset{m \in \mathcal{I}}{\overset{}{\bigcup}}AP_m$.

Following the terminology presented in Section \ref{sec:PaN}, the problem becomes:
\begin{prob} \thlabel{problem}
Synthesize a sequence of individual timed runs $r^t_1,...,r^t_N$ such that the following holds:
\begin{equation}\label{eq:problemdef2}
\left(r_G \models \phi_G\right)\wedge \left(r^t_1 \models \phi_1\wedge...\wedge r^t_N \models \phi_N\right)
\end{equation}
where the collective run $r_G$ was defined in \thref{def:Grun}.	
\end{prob}

\begin{rem}
Initially it might seem that if a run $r_G$ that satisfies the conjunction of the local formulas i.e., $r_G \models r_1^t \wedge \ldots \wedge r_N^t$ can be found, then the Problem \ref{problem} is solved in a straightforward centralized way. This does not hold since by taking into account the counterexample in \citep[Section III]{X2}, the following holds:
\begin{equation}
r_G^t \models  \bigwedge_{k \in \mathcal{I}}{\varphi_k} \nLeftrightarrow r_1^t \models \varphi_1 \wedge \ldots \wedge r_N^t \models \varphi_N. \label{eq: remark_import}
\end{equation}
\end{rem}

\section{Proposed Solution}\label{sec:sol}
The solution approach involves the following steps:
\begin{enumerate}
\item For each agent, we abstract the continuous-time linear system \eqref{eq:multi-agent dyn} into a WTS which describes the possible movements of the agent considering the dynamics and limitations of the state space (section \ref{sec:WTS}).
\item For each agent, we construct a local BWTS out of its WTS and a TBA representing the local MITL specification. The accepting timed runs of the local BWTS satisfy the local specification (section \ref{sec:lBWTS}).
\item Next, we construct a product BWTS out of the local BWTSs. The accepting timed runs of the product BWTS satisfy all local specifications (section \ref{sec:pBWTS}).
\item Next, we construct a global BWTS out of the product BWTS and the TBA representing the global MITL specification. The accepting runs of the global BWTS satisfy both the global specification and all local specifications (section \ref{sec:gBWTS}).
\item Finally, we determine the control input by applying a graph-search algorithm to find an accepting run of the global BWTS and projecting this accepting run onto the individual WTSs (section \ref{sec:design}).
\end{enumerate}

The computational complexity of the proposed approach is discussed in Section \ref{sec:complexity}.

\subsection{Constructing a WTS}\label{sec:WTS} 
In this section we consider the abstraction of the environment into a WTS. The definition of a WTS was given in Section \ref{sec:PaN}. The abstraction is performed for each agent $m \in \mathcal{I}$, resulting in $N$ number of WTSs.

Following the idea of \citep{A2}, we begin by dividing the state space $X_m$ into $p$-dimensional rectangles, defined as in \thref{def:rec}
\begin{defn}\thlabel{def:rec} 
A \textit{$p$-dimensional rectangle} $R_p(a,b)\subset\mathbb{R}^p$ is characterized by two vectors $a,b$, where $a=(a_1,a_2,...,a_p)$, $b=(b_1,b_2,...,b_p)$ and $a_i<b_i$, $\forall \: i=1,2,...,p$. The rectangle is then given by
\begin{equation}
R_p(a,b) = \{x\in\mathbb{R}^p : a_i\leq x_i\leq b_i, \forall i\in\{1,2,..,p\}\} \label{eq:req} 
\end{equation}
\end{defn} 
such that formula \eqref{eq:partition} is satisfied for each rectangle, i.e, such that each atomic proposition in the set $AP_m$ is either true at all points within a rectangle $R_p(a,b)$ or false at all points within the rectangle, i.e.
\begin{align}
ap_i &= (\top,\: \forall x\in R_p(a,b)) \ \text{or} \notag \\ 
ap_i &= (\perp,\: \forall x\in R_p(a,b)), \forall ap_i \in AP_m.
\label{eq:partition} 
\end{align} 
The set of states $\Pi=\{r_0,r_1,...,r_M\}$ of the WTS is then defined as the set of rectangles\\ $\mathcal{R}=\{R_p(a^0,b^0),R_p(a^1,b^1),...,R_p(a^M,b^M)\}$. From this, the definition of the initial state $\Pi_{init}$, transitions $\rightarrow$ and labelling $L$ follows directly:
\begin{equation}
\Pi_{init}=\{r_i\in \Pi| x_m^0\in R_p(a^i,b^i)\}
\end{equation}
\begin{eqnarray}
r_i\rightarrow r_j & \text{iff }R_p(a^i,b^i) \text{ and } R_p(a^j,b^j)\\
& \text{ have a common edge,}\nonumber
\end{eqnarray} 
\begin{equation}
L(r_i)=\{ap_i\in AP_m|ap_i=True \enspace \forall x\in R_p(a^i,b^i)\}
\end{equation}
 The set $\Sigma$ is given as the set of control inputs which induce transitions. In particular, a control input must be defined for each possible transition such that it guarantees the transition, that is no other transition can be allowed to occur and the edge of which the transition goes through must be reachable. This conditions on control inputs are required both to ensure that the synthesized path is followed and to guarantee that the following time estimation holds. A suggested low-level controller for a transition $r_k\rightarrow r_l$ in direction $i$, based on \citep{A2}, is given by
\begin{equation}
\begin{aligned}
& \underset{u_m \in U_m}{\text{max}}
& & [\dot{x}_m]_i \\
& \text{s. t.}
& & \left[\dot{x}_m\right]_i\geq \epsilon > 0, \\
& & & \left[\dot{x}_m\right]_j\leq -\epsilon, \forall j\neq i, j=\{1,...,p\}, \text{if} \ [x_m]_j=b^k_j, \\
& & & \left[\dot{x}_m\right]_j\geq\epsilon, \forall j\neq i, j=\{1,...,p\}, \text{if} \ [x_m]_j=a^k_j.
\end{aligned}
\end{equation}
 where $U_m=[-u_{max},u_{max}]$ is some bound on $u_m$ and $\epsilon$ is a robustness parameter. The idea is to maximize the transition speed, under the conditions that the speed in direction $j$ is negative at the edge with norm direction $j$, where $j$ is not the transition direction. 

Finally, the weights $d$ are assigned as the maximum transition times. These times are given according to \thref{eq:TFlinear} below. The theorem depends on the assumption $B_mu_m=B_{m1}x_m+B_{m2}$, where $B_{m1}$ and $B_{m2}$ are matrices of dimension $N\times N$ and $N\times 1$ respectively. The assumption corresponds to $u_m$ being affine.

\begin{thm}\thlabel{eq:TFlinear}
The maximum time $T^{max}(r_k,r_l)$ required for the transition $r_k\rightarrow r_l$ to occur, where $R_p(a^k,b^k)$ and $R_p(a^l,b^l)$ share the edge $e_{kl}$, $\overline{e_{kl}}$ is the edge located opposite to $e_{kl}$ in $R_p(a^k,b^k)$, $i$ is the direction of the transition, and assuming that $e_{kl}$ is reachable from all points within $r_k$, is defined as:
\begin{equation}
T^{max}(r_k,r_l)=\ln\left(\frac{([A_m^*]_{ii}x^0+C_m^*+[B_m^*]_i)}
{([A_m^*]_{ii}x^1+C_m^*+[B_m^*]_i)}\right)\frac{1}{[A_m^*]_{ii}}
\end{equation} 
where
\begin{eqnarray*}
C_m^*=\sum_{\underset{j\neq i}{j=1}}^n\min_{[x_m]_j\in r_k}\left([A_m^*]_{ij}
\left[x_m\right]_j\right),
\end{eqnarray*}
\begin{equation}
\min_{[x_m]_j\in r_k}\left([A_m^*]_{ij}[x_m]_j\right)=\left\lbrace\begin{array}{cc}
[A_m^*]_{ij}a_j^k & \text{if }[A_m^*]_{ij}>0\\
\left[A_m^*\right]_{ij}b_j^k & \text{if }[A_m^*]_{ij}<0
\end{array}\right.,
\end{equation}

and $x^0\in \overline{e_{kl}}$, $x^1\in e_{kl}$ (note that $x^0, x^1$ are the $i$th coordinate of the initial and final positions of the transition), $A_m^*=A_m+B_{m1}$ and $B_m^*=B_{m2}$, where $\dot{x}_m=A_mx_m+B_mu_m=A_mx_m+B_{m1}x_m+B_{m2}$.
\end{thm}

\begin{figure}
\centering
\includegraphics[width=0.4\textwidth]{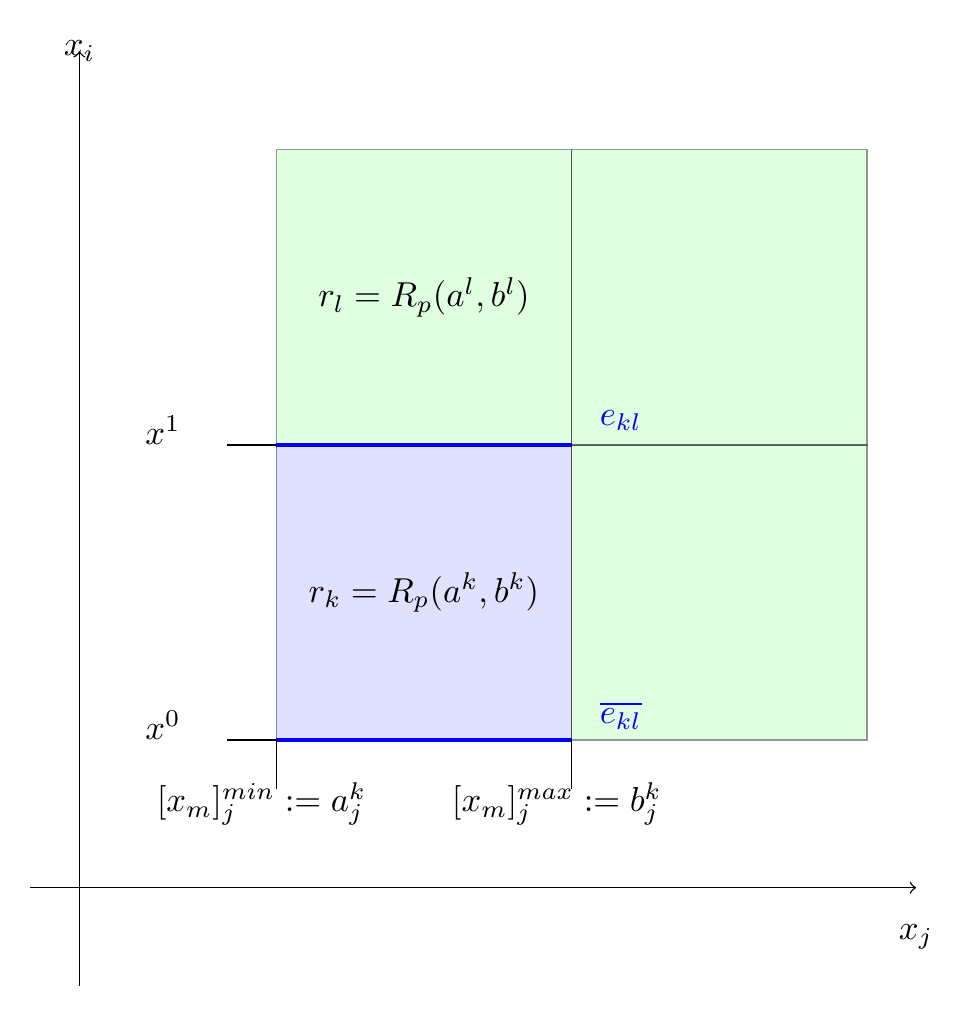}
\caption{Illustration of the variables in \thref{eq:TFlinear} in 2 dimensions.}\label{fig:Tmax}
\end{figure}

See figure \ref{fig:Tmax} for illustration of the variables of \thref{eq:TFlinear} in 2 dimensions.
 
\begin{pf}\textbf{of Theorem 1}\\ $T^{max}$ - the maximum transition time for $r_k\rightarrow r_l$ in a system following the linear dynamics \eqref{eq:multi-agent dyn} is determined by considering the minimum transition speed. Consider the dynamics of agent $m$ projected onto the direction of the transition $i$, i.e
\begin{align}
[\dot{x}_m]_i &= [A_mx_m+B_mu_m]_i, \label{eq:linearsystem} \\
\left[x_m(0)\right]_i &= [x_m^0]_i= x^0, \nonumber \\
\left[x_m(t_1)\right]_i &= [x_m^1]_i= x^1, \nonumber
\end{align}
where $x^0$ is the $i$th coordinate of some point on the edge $\overline{e}_{kl}$, and $x^1$ is the $i$th coordinate of some point on the edge $e_{kl}$. Since $B_mu_m=B_{m1}x_m+B_{m2}$, system \eqref{eq:linearsystem} can be rewritten to  \eqref{eq:linearsystem3}, by introducing $A_m^*=A_m+B_{m1}$ and $B_m^*=B_{m2}$.
\vspace{-6.0mm}
\begin{eqnarray}\label{eq:linearsystem3}
[\dot{x}_m]_i&=&[A_m^*]_{ii}[x_m]_i+\sum_{\underset{j\neq i}{j=1}}^n[A_m^*]_{ij}
\left[x_m\right]_j+[B_m^*]_i \\
&&\left[x_m(0)\right]_i=x^0\nonumber\\
&&\left[x_m(t_1)\right]_i=x^1\nonumber
\end{eqnarray}
The maximum transition time is determined by solving \eqref{eq:linearsystem3} for $t_1$. The equation can be solved by separating $[x_m]_i$ from $t$, if and only if $[A_m^*]_{ij}[x_m]_j$ is a constant $\forall j$. Since $[A_m^*]_{ij}$ is a constant this holds if and only if $[\dot{x}_m]_j=0$ or $[A_m^*]_{ij}=0$. Otherwise, the maximum transition time can be overestimated by considering the minimum transition speed $[\dot{x}_m]_i^{min}$, at every point in $r_k$, which can be determined by considering the limits of $[x_m]_j$ in $r_k$, namely $a_j^k$ and $b_j^k$
\begin{equation}
\min_{[x_m]_j\in r_k}\left([A_m^*]_{ij}[x_m]_j\right)=\left\lbrace\begin{array}{cc}
[A_m^*]_{ij}a_j^k & \text{if }[A_m^*]_{ij}>0\\
\left[A_m^*\right]_{ij}b_j^k & \text{if }[A_m^*]_{ij}<0
\end{array}\right.
\end{equation}
The maximum transition time, denoted $T^{max}$,  can then be overestimated as the solution to
\begin{eqnarray}
\dot{y}=[A_m^*]_{ii}y+C_m^*+[B_m^*]_i\\
y(0)=x^0\nonumber \\
y(T^{max})=x^1\nonumber
\end{eqnarray}
where $$C_m^*=\min_{\left[x_m\right]_j\in r_k}\sum\limits_{\underset{j\neq i}{j=1}}^{n}[A_m^*]_{ij}
\left[x_m\right]_j=\sum\limits_{\underset{j\neq i}{j=1}}^{n}\min_{[x_m]_j\in r_k}[A_m^*]_{ij}
\left[x_m\right]_j$$.
Which can be solved as: 
\begin{eqnarray}\label{eq:systemsolution}
\frac{dy}{dt}&=& [A_m^*]_{ii}y+C_m^*+[B_m^*]_i \implies\nonumber\\
\int dt&=&\int \left(\frac{1}{[A_m^*]_{ii}y+C_m^*+[B_m^*]_i}\right)dy\implies \nonumber\\
t+t_c&=&\frac{\ln([A_m^*]_{ii}y+C_m^*+[B_m^*]_i)}{[A_m^*]_{ii}}
\end{eqnarray}
Now, $y(0)=x^0$ yields
\begin{equation}\label{eq:k}
t_c=\frac{\ln([A_m^*]_{ii}x^0+C_m^*+[B_m^*]_i)}{[A_m^*]_{ii}}
\end{equation}
and $y(T^{max})=x^1$ yields
\begin{equation}\label{eq:t1b}
T^{max}=\ln\left(\frac{[A_m^*]_{ii}x^1+C_m^*+[B_m^*]_i}{[A_m^*]_{ii}x^0+C_m^*+[B_m^*]_i}\right)\frac{1}{[A_m^*]_{ii}}
\end{equation}

\end{pf}

\begin{rem}
If $C^*_m=0$ or $[\dot{x}_m]_j=0$ $\forall j$, then $T^{max}$ is the maximal time required for the transition to occur. Otherwise $T^{max}$ is an over-approximation.
\end{rem}
Finally, the weights of the WTS are defined as
\begin{eqnarray}
d(r_i,r_j) = T^{max}(r_i,r_j) \text{ where } (r_i, \sigma, r_j) \in \rightarrow.
\end{eqnarray}
for $\sigma=u_m(r_i,r_j)$.

\subsection{Constructing a Local BWTS}\label{sec:lBWTS}
Next, a local BWTS is constructed out of the WTS and a TBA representing the local MITL specification for each agent. As stated in Section \ref{sec:PaN} any MITL formula can be represented by a TBA \citep{MITL1}. Approaches for the translation were suggested in \citep{MITL2TA2}, \citep{MITL2TA5} and \citep{MITL2TA4}. Note that the time-intervals considered by the MITL formulas must be on the form $\leq a$ due to the over-approximation of time in the abstraction. The local BWTS is defined as:

\begin{defn}\thlabel{def:BWTS}
Given a weighted transition system $T=(\Pi, \Pi_{init},\Sigma, \rightarrow, AP, L,d)$ and a timed Büchi automaton $A=(S,S_{init},\mathcal{X},I,E,F,AP,\mathcal{L})$ their \textit{local BWTS} is defined as $T^p=T \otimes A=(Q,Q^{init},\leadsto, d^p,F^p, AP, L^p,I^p,C)$ with:
\begin{itemize}
\item $Q\subseteq\{(r,s)\in \Pi\times S:L(r)=\mathcal{L}(s)\}$,
\item $Q^{init}=\Pi_{init}\times S_{init}$
\item $q\leadsto q'$ iff
\begin{itemize}
\item $q=(r,s)$, $q'=(r',s')\in Q$
\item $(r,r') \in \rightarrow$ and
\item $\exists\: \gamma,R$, s.t.  $(s,\gamma,R,s')\in E$,
\end{itemize}
\item $d^p(q,q')=d(r,r')$ if $(q, q') \in \leadsto$,
\item $F^p=\{(r,s)\in Q:s\in F\}$ and
\item $L^p(r,s)=L(r)$
\item $I^p(q)=I(s)$
\item $C=\{c_1,...,c_M\}$\\
$c_i=\{(q, q')\enspace |\enspace \exists\enspace R \enspace s.t.\enspace (s,R,s')\in E$ and $x_i\in R \}$
\end{itemize}
where $M=\abs{\mathcal{X}}$.
\end{defn}

It follows from the construction and automata-based LTL model checking theory \citep{LTL2} that all possible runs of the local BWTS correspond to a possible run of the WTS. Furthermore, all accepting states of the local BWTS corresponds to accepting states in the TBA. This is formalized in \thref{lem1}.

\begin{lem}\thlabel{lem1}
An accepting timed run\\ $r^t_k=(q_k(0),\tau_k(0))(q_k(1),\tau_k(1))...$ of the local BWTS projects onto the timed run $r^t=(r(0),\tau(0))(r(1),\tau(1))...$ of the WTS that produces the timed word\\ $w(r^t)=(L_k(r(0)),\tau(0))(L_k(r(1)),\tau(1))...$ accepted by the TBA. Also, if there is a timed run that produces an accepting timed word of the TBA, then there is an accepting timed run of the local BWTS.
\end{lem}

\subsection{Constructing a Product BWTS}\label{sec:pBWTS}
Now, a product BWTS should be constructed from the local BWTSs. The definition is given as follows:

\begin{defn}\thlabel{def:PB}
Given $N$ local BWTSs $T^p_1,...,T^p_N$, defined as in \thref{def:BWTS}, and $M_k=\abs{\mathcal{X}_k}$ for $k\in \{1,..,N\}$, the\textit{ product BWTS} \begin{eqnarray*}
T_G&=& T^p_1\otimes...\otimes T^p_N=\\
&=&(Q_1,Q_1^{init},\leadsto_1, d^p_1,F^p_1, AP_1, L^p_1,I_1^p,C_1,M_1)\otimes...\\
&...&\otimes (Q_N,Q_N^{init},\leadsto_N, d^p_N,F^p_N, AP_N, L_N^p,I_N^p,C_N,M_N)\\ &=&(Q_G,Q_G^{init},\rightarrow_G,d_G,F_G,AP_G,L_G,I_G,C_G,M)
\end{eqnarray*}
is defined as:
\begin{itemize}
\item $Q_G\subseteq Q_1\times...\times Q_N$
\item $Q_G^{init}=Q_1^{init}\times...\times Q_N^{init}$
\item $(q_G, q'_G) \in \rightarrow_G$ iff
\begin{itemize}
\item $q_G=(q_1,...,q_N)\in Q_G$,
\item $q'_G=(q'_1,...,q'_N)\in Q_G$,
\item $\exists q'_k\in Q_k$ s.t. $(q_k, q'_k) \in \leadsto_k, \ \forall \ k \in \mathcal{I},$
\end{itemize}
\item $d_G(q_G,q'_G)=d_{max} = \max\limits_{i=\mathcal{I}}(d^p_i), \text{if} \ (q_G, q'_G) \in \rightarrow_G,$
\item $F_G=\{(q_1,...,q_N)\in Q_G \text{ s.t. } q_k\in F^p_k, \forall k\in \mathcal{I}\}$,
\item $AP_G=\bigcup\limits_{k \in \mathcal{I}}AP_k,$
\item $L_G(q_1,..,q_N)=\bigcup\limits_{k \in \mathcal{I}}L^p_k(q_k),$
\item $I_G(q_G) =\bigcup\limits_{k \in \mathcal{I} } I_k^p(q_k),$
\item $C_G=\{C^1,...,C^N\}$, $C^i=\{c^i_1,...,c^i_{M_i}\}$\\
$c^i_k=\begin{array}{l}
\{(q_G,q_G'), q_G=(q_1,...,q_N), q'_G=(q'_1,...,q'_N)\\ s.t\enspace
 (q_i,q_i') \in c_k, c_k\in C_i\}
\end{array}$
\item $M=\{M_1,..,M_N\}$
\end{itemize}
\end{defn}

It follows from the construction that an accepting collective run of the product BWTS corresponds to accepting runs of each local BWTS. Formally

\begin{lem}\thlabel{lem2}
An accepting collective run $r_G$ of the product BWTS projects onto an accepting timed run $r^t_k$ of a local BWTS, for each $k\in I$. Moreover, if there exists an accepting timed run for every local BWTS, then there exists an accepting collective run.
\end{lem}

\begin{rem}
Note that the definition does not allow for the agents to start transitions at different times. This causes an overestimation of required time which increases the risk for false negative result. An alternative definition which allows the mentioned behaviour was suggested in \citep{X2}. However, the definition suggested here requires 
less number of states and hence less computational time. 
\end{rem}

\subsection{Constructing a Global BWTS}\label{sec:gBWTS}
Finally, a global BWTS is constructed from the product BWTS and a TBA representing the global MITL specification.

\begin{defn}\thlabel{def:gBWTS}
Given a product BWTS\\ $T_G=(Q_G,Q_G^{init},\rightarrow_G,d_G,F_G,AP_G,L_G,I_G,C_G,M)$ and a global TBA $A_G=(S_G,S_G^{init},\mathcal{X}_G,I_G,E_G,\mathcal{F}_G,\mathcal{L}_G)$, with $M_G=\abs{\mathcal{X}_G}$, their \textit{global BWTS} $\hat{T}_G=T_G\otimes A_G=(\hat{Q}_G,\hat{Q}_G^{init},\leadsto_G,\hat{d}_G,\hat{F}_G,AP_G,\hat{L}_G)$ is defined as:
\begin{itemize}
\item $\hat{Q}_G\subseteq \{(q,s)\in Q_G\times S_G$ s.t. $L_G(q)=\mathcal{L}_G(s)\}\times Z_0\times...\times Z_N\times \{1,2\}$, where $Z_i=\{z_1^i,...,z_{M_i}^i\}$ for $i=1,...,N$ and $Z_0=\{z_1^0,...,z_{M_G}^0\}$
\item $\hat{Q}_G^{init}=Q_G^{init}\times S_G^{init}\times \{1,..,1\}\times...\times \{1,...,1\}\times \{1,2\}$, where $\{1,...,1\}\times...\times \{1,...,1\}$ consists of $N+1$ sets, where the first set contains $M_G$ ones, and the remaining sets contains $M_i$ ones each,
\item $(q_G, q'_G) \in \leadsto_G$ iff
\begin{itemize}
\item $q_G=(q,s,Z_0,...,Z_{N},l)\in \hat{Q}_G$,
\item $q_G'=(q',s',Z'_0,...,Z'_{N},l')\in \hat{Q}_G$,
\item $(q, q') \in \rightarrow_G$,
\item $\exists \gamma, R$ s.t. $(s,\gamma,R,s')\in E_G$ s.t,
\begin{itemize}
\item For all $i\in \{1,...,N\}$, $Z_i$ and $Z'_i$ are such that $$\begin{array}{ll}
z_k^i=0 \text{ and } z_k^{i'}=1,&\text{if } (q,q')\in c_k^i\\
z_k^{i'}=z_k^i, & \text{otherwise}
\end{array} $$
\item $Z_0$ and $Z_0'$ are such that 
$$z_k^0=\left\lbrace\begin{array}{ll}
0& \text{if } x_k\in R\\
1& \text{otherwise}
\end{array}\right.$$
$$z_k^{0'}=\left\lbrace\begin{array}{ll}
1& \text{if } x_k\in R\\
z_k^0& \text{otherwise}
\end{array}\right.$$
\end{itemize}
\item $l'=\left\lbrace\begin{array}{ll}
1,&\text{if }l=1 \text{ and } q\in F_G\\
& \text{or }l=2 \text{ and } s\in \mathcal{F}_G\\
2,&\text{otherwise}
\end{array}\right. $ 
\end{itemize}
\item $\hat{d}_G(q_G,q'_G)=d_G(q,q')$ if $(q_G, q_G') \in \leadsto_G$,
\item $\hat{F}_G=\{(q,s,Z_0,...,Z_{N},1)\in \hat{Q}_G$ s.t. $q\in F_G\}$ and
\item $\hat{L}_G(q,s,Z_0,...,Z_{N},l)=L_G(r)$.
\item $I(q_G)=I_G(q)\cup I(s)$
\end{itemize}
\end{defn}

It follows from the construction that an accepting run of the global BWTS corresponds to an accepting run of the product BWTS as well as an accepting run of the TBA representing the global specification. Formally

\begin{lem}\thlabel{lem3}
An accepting timed run $r_G^t$ of the global BWTS projects onto an accepting collective run $r_G$ of the product BWTS that produces a timed word $w(r_G)$ which is accepted by the TBA representing the global specification. Also, if there exists an accepting collective run that produces a timed word accepted by the TBA, then there is an accepting timed run $r_G^t$ of the global BWTS.
\end{lem}

\subsection{Control Synthesis}\label{sec:design}
The controller can now be designed by applying a modified graph-search algorithm (such as a modified Dijkstra) to find an accepting run of the global product. The modification of the algorithm includes a clock valuation when considering a transition. A sketch of the modification is given in Algorithm \ref{psuedo-code}. The idea is to calculate the clock valuation for each clock given the predecessors of the current state, if a valuation does not satisfy the clock constraint the transition is not valid. When the algorithm is complete the accepting run is projected onto the WTSs following \thref{lem1}, \thref{lem2} and \thref{lem3}. Finally, the set of controllers are given as the sequences of control inputs which induces the timed runs $(r^t_1, r^t_2,...r^t_N)$ which in turn produce accepted timed words of all local TBAs as well as of the global TBA.

\begin{algorithm2e}
\KwResult{Clock Valuation}
 $M$=Total number of clocks\;
 $q$=current state; $q'$=possible successor of $q$\;
 \For{$i=1:M$}{
 $v_i=d(q,q')$;
 $k=q$\;
 \If{$z_i(k)==1$}{
 \While{$z_i(Pred(k))==1$}{
  $v_i$=$v_i+d(Pred(k),k)$\;
  $k=Pred(k)$\;
  \If{$Pred(k)$ isempty}{break;}}
 }}
  
 \If{$v_1,...,v_M\nvDash I(q)$} {Transition is illegal - don't add $q$ as a successor to $Pred(q)$.}
 \caption{Modification to search-algorithm to evaluate clocks}\label{psuedo-code}
\end{algorithm2e}

\subsection{Complexity} \label{sec:complexity}
The framework proposed in this paper requires at most
\begin{equation}
|\hat{T}_G|=\prod_{i=1}^N (|T_i|\times |A_i|\times 2^{M_i})\times |A_G|\times 2^{M_G} \times 2
\end{equation}
number of states. The method suggested in \citep{X2} requires
\begin{align*}
|\hat{T}_G|=&\prod_{i=1}^N \left(|T_i|\times |A_i|\times (C_{max,i}+1)^{M_i}\right)\times\\ 
&\times |A_G|\times 2\times (C_{max,G}+1)^{M_G}\times 2
\end{align*}
number of states, where all possible clock values are integers in the set $[0,C_{max,i}]$ and $[0,C_{max,G}]$ for the local and global TBA's respectively. Hence the number of states required in the proposed framework is a factor 
$$\dfrac{\prod_{i=1}^N (C_{max,i}+1)^{M_i}\times (C_{max,G}+1)^{M_G}}{2^{\sum_{i=1}^N (M_i)+M_G}}$$ less.

\section{Simulation Result}\label{sec:ex} 
\begin{figure}
\centering
\captionsetup{width=0.4\textwidth}
\includegraphics[width=0.3\textwidth]{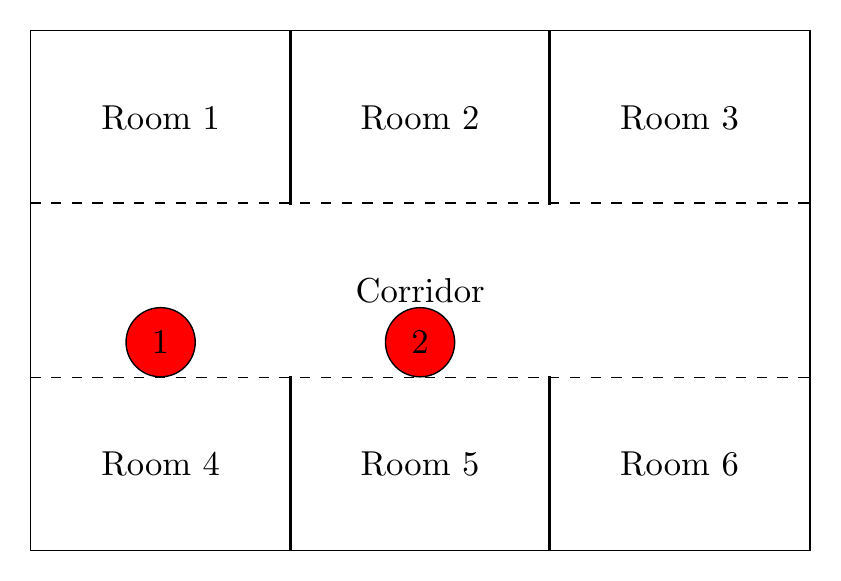}
\caption{Draft of the problem described in section \ref{sec:ex}.}
\label{fig:draft}
\end{figure}

Consider $N=2$ agents with dynamics in the form:
\begin{subequations}
\begin{align}
\dot{x} &= \begin{bmatrix}
2 & 1 \\
0 & 2
\end{bmatrix} x
+
\begin{bmatrix}
1 & 0 \\
0 & 1
\end{bmatrix} u \\
\dot{x} &= 
\begin{bmatrix}
1 & 0 \\
0 & 1
\end{bmatrix} x 
+
\begin{bmatrix}
0 & 1 \\
1 & 0
\end{bmatrix} u
\end{align}
\end{subequations}
evolving in a bounded workspace $X$ consisting of 6 rooms and a hallway as can be seen in Figure \ref{fig:draft}. Each agent is assigned with the local MITL formula $\phi_L=\lozenge_{0.1}r_2 \wedge r_2\rightarrow \lozenge_{0.3} r_6$ ('Eventually, within $0.1$ time units, the agent must be in room 2, and if the agent enters room 2 it must then enter room 6 within $0.3$ time units.'). Furthermore, they are assigned with the global MITL formula $\phi_G=\lozenge_{1}(a_1=r_1 \wedge a_2=r_2)$ ('Eventually, within $1$ time units, agent 1 must be in room 1 and agent 2 must be in room 2, at the same time.'). The initial positions of each agent is indicated by the encircled 1 and 2 in Figure \ref{fig:draft}.
\begin{figure}
\centering
\captionsetup{width=0.4\textwidth}
\includegraphics[width=0.4\textwidth]{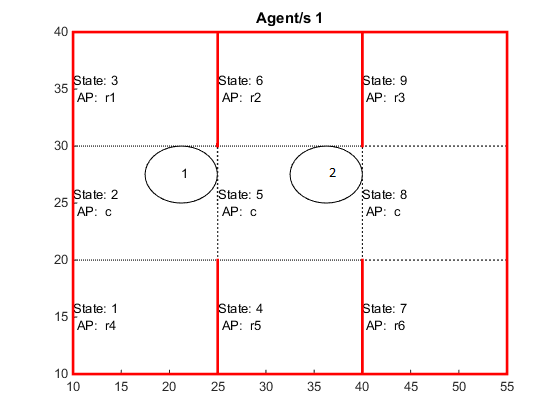}
\caption{Partition constructed by the MATLAB script. The circles represents the initial states of each agent.}
\label{fig:part:MATLAB}
\end{figure}
\begin{rem}
As can be seen in figure \ref{fig:draft}, some walls have been added to the environment. Transitions through these are forbidden. This is handled by the abstraction since the edges on which the walls are placed aren't reachable.
\end{rem}
The suggested environment can be abstracted to a WTS of 9 states (see figure \ref{fig:part:MATLAB}), while the local MITL formula can be represented by a TBA of 4 states. This results in a local BWTS of 36 states. Notable is that the local BWTSs for each agent will be identical if and only if the dynamics are identical. Furthermore, if the problem at hand only considers local MITL formulas - that is, if no global tasks are considered - the five step procedure described earlier can stop here. In that case, the control design can be performed based on accepting runs of each local BWTS. Since a global task is considered in this case, the product BWTS and the global BWTS must be constructed. The product BWTS will consist of  $(|Q_1|\cdot |Q_2|)=1296$ states while the global BWTS will consist of $2\cdot(|Q_{pBTWS}|\times |Q_{gTBA}|\times 2^{M_1}\times 2^{M_2}\times 2^{M_G})=248832$ states. MATLAB was used to simulate the problem by constructing all transition systems and applying a modified Dijkstra algorithm to find an accepting path as well as a control sequence that satisfies the specifications. 

\begin{figure}
\centering
\captionsetup{width=0.5\textwidth}
\begin{subfigure}[t]{0.59\textwidth}
\centering\captionsetup{width=0.9\textwidth}
\includegraphics[width=\textwidth]{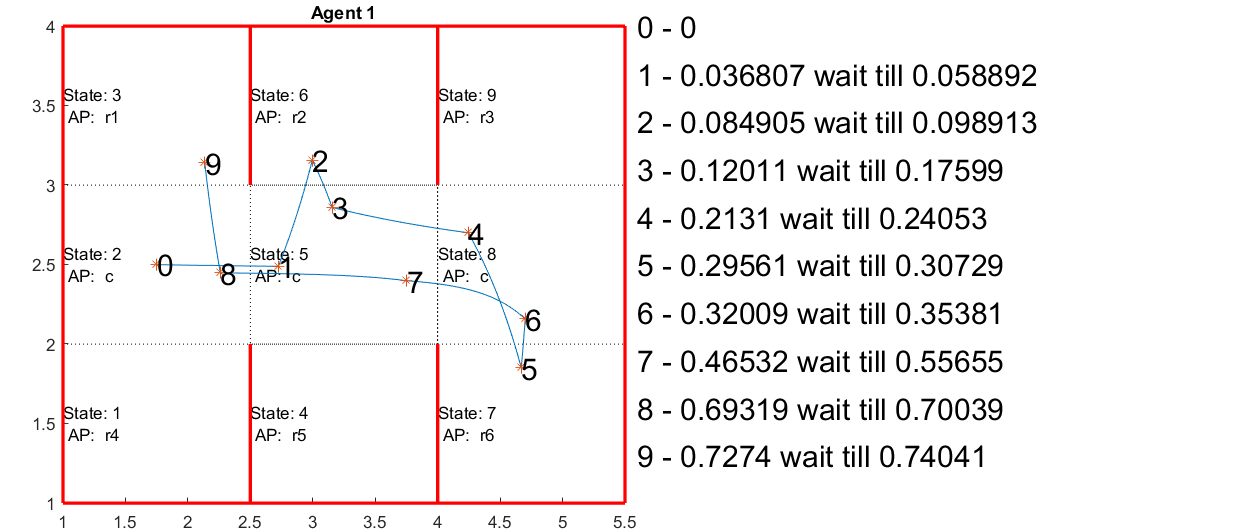}
\caption{Agent 1}
\end{subfigure}
\begin{subfigure}[t]{0.59\textwidth}
\centering\captionsetup{width=0.9\textwidth}
\includegraphics[width=\textwidth]{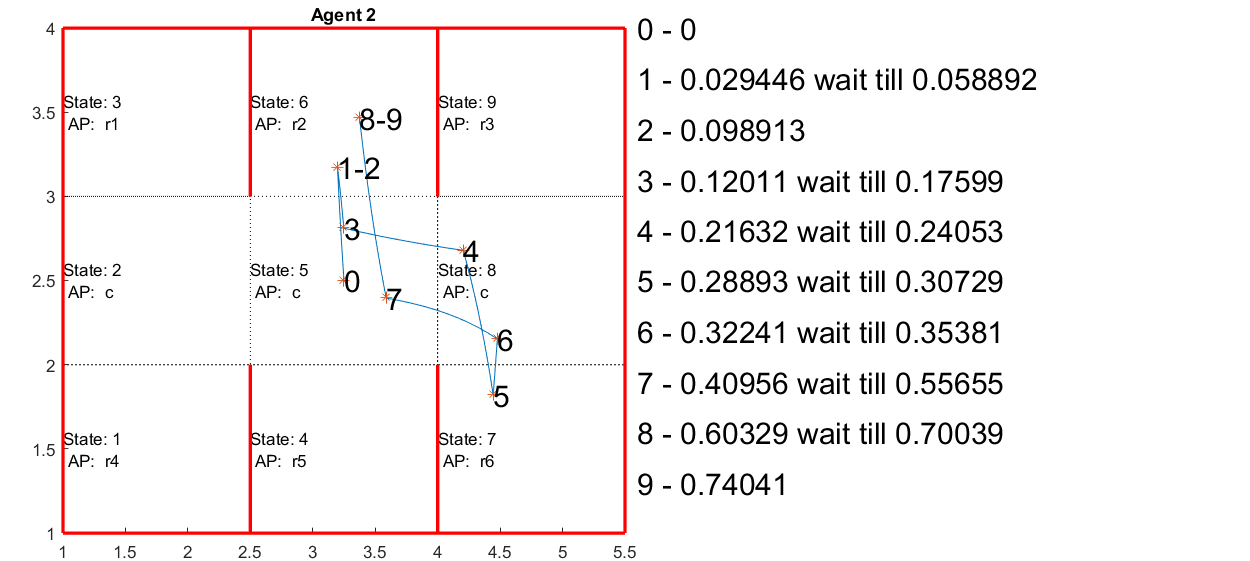}
\caption{Agent 2}
\end{subfigure}
\caption{Illustration of the paths of each agent in the example. The numbers 0-9 represent the end of each joined transition. The actual arrival time at each location as well as the time the agent is required to wait till the worst case transition time has been reached (and it is guaranteed that all other agents have transitioned), is noted to the right of the figure. The time the agent has to wait till corresponds to the worst case estimation of the required transition time and is due to the requirement that the agents make transitions simultaneously. It is notable that both agents finish all transitions on less time than the worst case estimation. Hence, the waiting time can be further cut by allowing the agents to communicate to each other when a transition is done.}
\label{fig:path}
\end{figure}

The projection of the found accepting run onto each WTS, yielded $[ 2,\: 5,\: 6,\: 5,\: 8,\: 7,\: 8,\: 5,\: 2,\:3 ]$ and $[ 5,\: 6,\: 6,\: 5,\: 8,\: 7,\: 8,\: 5,\: 6,\: 6 ]$, for the respective agent. The result is visualized in Figure \ref{fig:path}, which shows the evolution of each closed-loop system for the given initial positions. The figure was constructed by implementing the built-in function \textit{ode45} for the determined closed-loop system in each state with the initial position equal to the last position of the former transition. The switching between controllers is performed based on the position of the agent; namely the switching from controller $u_{ij}$ to $u_{jk}$ is performed when the agent has entered far enough into state $j$, where ''far enough'' was defined as 5 iterations of \textit{ode45} upon exiting the previous state. The estimated time distances for each joined transition are given in table \ref{tab:time}.
\begin{table}\centering
\captionsetup{width=0.4\textwidth}
\caption{The worst case estimation of the transition times as well as the actual required time. The actual transition times are defined as the maximum of the times the individual agents require to complete the transition. *These transitions require agent 2 to stay in place, hence the actual time is here defined as the time agent 1 requires to complete the transition.$\star$Numbered in order of transitions, see figure \ref{fig:path}.}\label{tab:time}
\begin{tabular}{p{1.0cm} p{1.0cm} p{1.0cm} p{2.2cm} p{0.8cm}}
Position$\star$& Agent 1& Agent 2& Worst Case Time Estimation& Actual Time\\
\hline
\\
0& 2& 5& 0& 0\\
1& 5& 6& 0.0589& 0.0368\\
2& 6& 6& 0.04& 0.026*\\
3& 5& 5& 0.0771& 0.0212\\
4& 8& 8& 0.0645& 0.0403\\
5& 7& 7& 0.0668& 0.0551\\
6& 8& 8& 0.0465& 0.0151\\
7& 5& 5& 0.2027& 0.1115\\
8& 2& 6& 0.1438& 0.1366\\
9& 3& 6& 0.04& 0.027*
\end{tabular} 
\end{table}
That is, the worst case transition times yields;\\  
\begin{tabular}{c p{6.5cm}}
0& Agent 1 and Agent 2 begins at their respective initial position in the corridor\\
1& Agent 2 enters room 2 within 0.0589 time units from start\\
2& Agent 1 enters room 2 within 0.0989 time units from start\\
5& Agent 1 and Agent 2 enters room 6 within 0.2084 and 0.2484 time units respectively from entering room 2\\
9& Agent 1 is in room 1 while Agent 2 is in room 2 within 0.7404 time units from start.
\end{tabular}

From this, it is clear that the given path will satisfy the MITL formulas. 
\newpage
The simulation presented in this section was run in MATLAB on a laptop with a Core i7-6600U 2.80 GHz processor, the runtime was approximately 30min.
\section{Conclusions and Future Work}\label{sec:conc}
A correct-by-construction framework to synthesize a controller for a multi-agent system following continuous linear dynamics such that some local MITL formulas as well as a global MITL formula are satisfied, has been presented. The method is supported by result of the simulations in the MATLAB environment. Future work includes communication constraints between the agents.

%

\bibliographystyle{unsrt}
\bibliography{ifacBib}

\begin{thebibliography}{19}
\providecommand{\natexlab}[1]{#1}
\providecommand{\url}[1]{\texttt{#1}}
\providecommand{\urlprefix}{URL }
\expandafter\ifx\csname urlstyle\endcsname\relax
  \providecommand{\doi}[1]{doi:\discretionary{}{}{}#1}\else
  \providecommand{\doi}{doi:\discretionary{}{}{}\begingroup
  \urlstyle{rm}\Url}\fi

\bibitem[{Alur and Dill(1994)}]{TA1}
Alur, R. and Dill, D.L. (1994).
\newblock A theory of timed automata.
\newblock \emph{Theoretical computer science}, 126(2), 183--235.

\bibitem[{Alur et~al.(1996)Alur, Feder, and Henzinger}]{MITL1}
Alur, R., Feder, T., and Henzinger, T.A. (1996).
\newblock The benefits of relaxing punctuality.
\newblock \emph{Journal of the ACM (JACM)}, 43(1), 116--146.

\bibitem[{Baier and Katoen(2007)}]{LTL2}
Baier, C. and Katoen, J.P. (2007).
\newblock \emph{Principles of model checking}.
\newblock MIT press.

\bibitem[{Brihaye et~al.(2013)Brihaye, Esti{\'e}venart, and
  Geeraerts}]{MITL2TA5}
Brihaye, T., Esti{\'e}venart, M., and Geeraerts, G. (2013).
\newblock On mitl and alternating timed automata.
\newblock In \emph{Formal Modeling and Analysis of Timed Systems}, 47--61.
  Springer.

\bibitem[{Fu and Topcu(2015)}]{MITL2TA}
Fu, J. and Topcu, U. (2015).
\newblock Computational methods for stochastic control with metric interval
  temporal logic specifications.
\newblock \emph{CoRR}, abs/1503.07193.

\bibitem[{Gol and Belta(2013)}]{A2}
Gol, E.A. and Belta, C. (2013).
\newblock Time-constrained temporal logic control of multi-affine systems.
\newblock \emph{Nonlinear Analysis: Hybrid Systems}, 10, 21--33.

\bibitem[{Guo and Dimarogonas(2015)}]{guo_2015_reconfiguration}
Guo, M. and Dimarogonas, D. (2015).
\newblock {M}ulti-{A}gent {P}lan {R}econfiguration {U}nder {L}ocal {LTL}
  {S}pecifications.
\newblock \emph{The International Journal of Robotics Research}, 34(2),
  218--235.

\bibitem[{Kantaros and Zavlanos(2016)}]{zavlanos_2016_multi-agent_LTL}
Kantaros, Y. and Zavlanos, M. (2016).
\newblock {A} {D}istributed {LTL}-{B}ased {A}pproach for {I}ntermittent
  {C}ommunication in {M}obile {R}obot {N}etworks.
\newblock \emph{American Control Conference (ACC), 2016}, 5557--5562.

\bibitem[{Karaman and Frazzoli(2008)}]{frazzoli_MTL}
Karaman, S. and Frazzoli, E. (2008).
\newblock {V}ehicle {R}outing {P}roblem with {M}etric {T}emporal {L}ogic
  {S}pecifications.
\newblock \emph{47th IEEE Conference on Decision and Control (CDC 2008)},
  3953--3958.

\bibitem[{Kloetzer et~al.(2011)Kloetzer, Ding, and
  Belta}]{belta_cdc_reduced_communication}
Kloetzer, M., Ding, X.C., and Belta, C. (2011).
\newblock {M}ulti-{R}obot {D}eployment from {LTL} {S}pecifications with
  {R}educed {C}ommunication.
\newblock \emph{50th IEEE Conference on Decision and Control (CDC 2011)},
  4867--4872.

\bibitem[{Kloetzer and Belta(2008)}]{A1}
Kloetzer, M. and Belta, C. (2008).
\newblock A fully automated framework for control of linear systems from
  temporal logic specifications.
\newblock \emph{Automatic Control, IEEE Transactions on}, 53(1), 287--297.

\bibitem[{Kress-Gazit et~al.(2007)Kress-Gazit, Fainekos, and Pappas}]{LTL1}
Kress-Gazit, H., Fainekos, G.E., and Pappas, G.J. (2007).
\newblock Where is waldo? sensor based temporal logic motion planning.
\newblock \emph{mag}.

\bibitem[{Loizou and Kyriakopoulos(2004)}]{loizou_2004}
Loizou, S. and Kyriakopoulos, K. (2004).
\newblock {A}utomatic {S}ynthesis of {M}ulti-{A}gent {M}otion {T}asks {B}ased
  on {LTL} {S}pecifications.
\newblock \emph{43rd IEEE Conference on Decision and Control (CDC 2004)}, 1,
  153--158.

\bibitem[{Maler et~al.(2006)Maler, Nickovic, and Pnueli}]{MITL2TA2}
Maler, O., Nickovic, D., and Pnueli, A. (2006).
\newblock From mitl to timed automata.
\newblock In \emph{Formal Modeling and Analysis of Timed Systems}, 274--289.
  Springer.

\bibitem[{Ni{\v{c}}kovi{\'c} and Piterman(2010)}]{MITL2TA4}
Ni{\v{c}}kovi{\'c}, D. and Piterman, N. (2010).
\newblock \emph{From MTL to deterministic timed automata}.
\newblock Springer.

\bibitem[{Nikou et~al.(2016{\natexlab{a}})Nikou, Boskos, Tumova, and
  Dimarogonas}]{alex_acc_2017}
Nikou, A., Boskos, D., Tumova, J., and Dimarogonas, D.V. (2016{\natexlab{a}}).
\newblock {C}ooperative {P}lanning for {C}oupled {M}ulti-{A}gent {S}ystems
  under {T}imed {T}emporal {S}pecifications.
\newblock \emph{http://arxiv.org/pdf/1603.05097v2.pdf}.

\bibitem[{Nikou et~al.(2016{\natexlab{b}})Nikou, Tumova, and Dimarogonas}]{X2}
Nikou, A., Tumova, J., and Dimarogonas, D.V. (2016{\natexlab{b}}).
\newblock Cooperative task planning of multi-agent systems under timed temporal
  specifications.

\bibitem[{Raman et~al.(2015)Raman, Donz{\'e}, Sadigh, Murray, and
  Seshia}]{murray_2015_stl}
Raman, V., Donz{\'e}, A., Sadigh, D., Murray, R., and Seshia, S. (2015).
\newblock {R}eactive {S}ynthesis from {S}ignal {T}emporal {L}ogic
  {S}pecifications.
\newblock \emph{18th International Conference on Hybrid Systems: Computation
  and Control (HSCC 2015)}, 239--248.

\bibitem[{Zhou et~al.(2016)Zhou, Maity, and Baras}]{baras_MTL_2016}
Zhou, Y., Maity, D., and Baras, J.S. (2016).
\newblock {T}imed {A}utomata {A}pproach for {M}otion {P}lanning {U}sing
  {M}etric {I}nterval {T}emporal {L}ogic.
\newblock \emph{European Control Conference (ECC 2016)}.

\end{thebibliography}
\end{document}